\documentstyle[epsfig]{aipproc}
\begin{document}
\title{Cosmological Parameter Estimation\\ from CMB Experiments
\footnote{To appear in {\it``Cosmology and Particle Physics''}, 
Proc.\ of the CAPP 2000 Conference, Verbier, Switzerland, July 2000, 
eds.\ J. Garc\'{\i}a-Bellido, R. Durrer and M. Shaposhnikov (AIP, 2001)} }

\author{Amedeo Balbi}
\address{Dipartimento di Fisica, Universit\`a Tor Vergata\\
Via della Ricerca Scientifica\\
Roma, I-00133, Italy}

%\lefthead{LEFT head}
%\righthead{RIGHT head}
\maketitle

\begin{abstract}
I review the general aspects of cosmological parameter estimation
from observations of the cosmic microwave background (CMB) temperature 
anisotropies in the framework of inflationary adiabatic models. 
The most recent CMB datasets are starting to give good constraints on 
the relevant parameters of inflationary adiabatic models. They point toward
a model consistent with the basic predictions of inflation: a nearly
flat universe, with a nearly scale invariant spectrum of primordial 
fluctuation.
\end{abstract}

\section*{Introduction}
The study of the cosmic microwave background (CMB) temperature anisotropies
has long been recognized as one of the most powerful tools to answer the
basic questions about the nature of the universe: what is its geometry, 
its matter and energy content, what are the initial conditions which seeded
the formation of structure, etc.
It is a firm theoretical conclusion that the angular distribution of
the CMB anisotropies must encode a vast amount of information on the
cosmological parameters. The majority of this information is thought
to be concentrated at angular scales smaller than about 1 degree on
the sky, corresponding to regions of the universe that were in casual
contact when the background photons decoupled from the matter (at
redshifts of about 1000). On this scales, physical processes in the
early universe were able to leave an imprint on the CMB.

The pioneering observations by the COBE satellite in the early 90's, 
which led to the first unambiguous detection of the CMB anisotropies 
\cite{balbi:smoot92}, 
were followed by a large number of ground based and balloon-borne 
observations which attempted to collect information on the fine-structure 
pattern of the anisotropy, where most of the dependence on cosmological 
parameters is encoded. 
Recently, the BOOMERanG
\cite{balbi:debe00} and MAXIMA \cite{balbi:hanany00}
balloon-borne experiments produced
the first high-resolution maps of the CMB anisotropy pattern, and
moved us closer to the goal of a long-awaited
high precision measurement of the CMB angular power spectrum.
These results have been used by several authors 
\cite{balbi:balbi00} \cite{balbi:lange00} \cite{balbi:tegmark00}
\cite{balbi:jaffe00} to obtain high-precision
constraints on the set of cosmological parameters which defines
the inflationary adiabatic class of models. 

In this review I will attempt to give an idea of how the 
cosmological parameters affect the CMB observables, and of 
the process which leads from the observation of the CMB anisotropy to 
the extraction of such parameters. The current constraints on 
inflationary models from the CMB will also be discussed.

\section*{The CMB anisotropy dependence on cosmological parameters}

It is generally believed  that the observed large scale structure of 
the Universe formed by gravitational amplification of small density 
perturbations generated in the early universe.  
In such {\em gravitational instability} scenarios, the presence of 
anisotropies in the temperature distribution of the CMB is unavoidable: 
density fluctuations must leave an imprint in the CMB at
the time of photon-matter decoupling, at redshifts of about 1000. 

The anisotropy as a function of the direction of observations can be
expanded in spherical harmonics:
\begin{equation}
{\delta T\over T}(\hat{\gamma})=
\sum_{\ell m} a_{\ell m} Y_{\ell m}(\theta,\phi).
\end{equation}

The coefficients $C_\ell \equiv\langle\vert a_{\ell m}\vert^2\rangle$
define the angular power spectrum of the CMB anisotropy.  
The $C_\ell$'s do not depend on the azimuthal index
$m$ as a consequence of the isotropy of space.  
For Gaussian initial conditions, the angular power spectrum
$C_\ell$ carries all the information on the angular temperature
anisotropy of the CMB. Each $\ell$ probes an angular scale $\theta$ 
on the sky given approximately by $\ell \sim \pi/\theta$. 
Only $\ell$'s corresponding to angular scales
which were in casual contact at decoupling may have been affected by physical
processes prior to decoupling. 
For this reason the dependence on physical parameters is mostly
found at high $\ell$'s (small angular scales), while low $\ell$'s probe 
the primordial shape of the power spectrum\footnote{Neglecting secondary
processes which may alter the spectrum after decoupling.}.

Within the inflationary adiabatic family, a
given cosmological model is specified by the value of a number of parameters.
These include the fractional density of matter in the
universe, which is the sum of contributions from baryons and cold dark matter, 
$\Omega_m\equiv \Omega_b+\Omega_{cdm}$; the fractional density of
vacuum energy, $\Omega_\Lambda$; the total energy density, 
$\Omega\equiv \Omega_m+\Omega_\Lambda$, which defines the curvature 
of the universe through $\Omega_k\equiv 1-\Omega$; 
the Hubble constant, parameterized by its value $h$ in units of 
100 km s$^{-1}$ Mpc$^{-1}$; 
and the amplitude $A$ and spectral index $n$ of the primordial power spectrum
of density fluctuations, modeled as $A k^n$ (but see \cite{balbi:schwarz00} 
for an alternative view).  
The $C_\ell$'s corresponding to a set of parameters can be computed exactly 
using high-accuracy numerical codes \cite{balbi:cmbfast}.  
A simplified analytical treatment was used by 
some authors (see e.g. \cite{balbi:hu} for an excellent review on the
subject) in order to give a better intuition of how different physical
processes leave an imprint on the CMB angular power spectrum.

Before decoupling, the photons and baryons are tightly coupled by
different scattering processes. The cold dark matter is non-interacting and
contributes only to the gravitational potential.  The dynamics of the
photon-baryon fluid is described by an equation reminiscent of the
classical Jeans equation governing the perturbation in a
self-gravitating gas, which for a given wave-number $k$ is:
\begin{equation}
\ddot{\delta} + {\dot{a}\over a}{R\over 1+R}\dot{\delta} +
k^2c_s^2\delta = F
\end{equation}
where $a$ is the scale factor describing the expansion of the
universe, $R\equiv 3\rho_B/4\rho_\gamma$ is the baryons to photons density
ratio and $c_s=c/[3(1+R)]^{1/2}$ is the
sound velocity.  For adiabatic initial conditions 
(i.e. conserving the entropy of the radiation per baryon) the intrinsic
temperature fluctuation of the photons, $\delta T/T$ is
related to the matter density perturbation simply by $\delta T/T=\delta/3$.  
The dots represent derivatives with respect to the conformal time
$\eta\equiv\int dt/a$. The term $F$ describes the
gravitational effects and can be held approximately
constant near decoupling. 
So, the evolution of the perturbations prior to decoupling 
is essentially governed by a forced harmonic
oscillator equation. The expansion of the universe introduces a viscosity 
term through $\dot{a}/a$ that may be neglected for the
sake of simplicity.

The physical interpretation of this equation is very simple. The baryons
tend to collapse due to self gravitation. The restoring force is
provided by the radiation pressure $k^2c^2/3$. This sets up acoustic
oscillations (the sound velocity $c_s$ quantifies the resistance of the
fluid to compression).  The higher $R$, the larger the amplitude of
the oscillations. The driving force term due to gravitation, constant
in our approximation, simply displaces the zero point of the
oscillations. Increasing $R$ (i.e. the baryon content of the universe)
enhances this displacement, and gives more amplitude to compressions 
over rarefactions, because of the increased inertia of the fluid. 
If we freeze the oscillations at the time of decoupling $\eta_{dec}$, 
each mode will be in a different stage of oscillation. 
The total power will have the largest
contributions from modes having $k c_s\eta_{dec}=k r_s^{dec} =m\pi$,
where $r_s^{dec}\equiv c_s\eta_{dec}$ defines the physical scale of the
sound horizon at decoupling.
This results in a harmonic series of peaks in the angular power spectrum, 
whose position is related to $\theta_s$, the characteristic angular 
scale subtended by the sound horizon at decoupling, by 
$\ell\sim \pi/\theta_s$. 
Odd peaks are due to compression of the fluid, even peaks to
rarefaction: so the odd peaks will be generally higher than the even
peaks because of $R$.  Increasing the baryon content will enhance this effect.

To calculate the dependence of the peaks position on cosmological parameters
we have to specify the angular diameter distance relation which maps a given
physical scale at decoupling into an angle $\theta$. As we saw, 
the relevant physical scale is the sound horizon at decoupling 
$r_s^{dec}= c_s\eta_{dec}$. The angular scale of the sound horizon at
decoupling, $\theta_s$, is approximately given by:
\begin{equation}
\theta_s^{-1} \approx {\sinh
I\over 2c_s~a_{dec}^{1/2} }
\left({\Omega_m\over \Omega_k}\right)^{1/2};\;\; 
I=\int_{a_{dec}}^1 {c~\Omega_k^{1/2} da\over 
 [\Omega_m a+\Omega_\Lambda a^4+\Omega_k a^2]^{1/2}},
\end{equation}
for an open universe, $\Omega_k > 0$;
\begin{equation}
\theta_s^{-1} \approx {\sin I\over 2c_s~a_{dec}^{1/2} }
\left({\Omega_m\over -\Omega_k}\right)^{1/2};\;\; 
I=\int_{a_{dec}}^1 {c~(-\Omega_k)^{1/2} da\over 
 [\Omega_m a+\Omega_\Lambda a^4+\Omega_k a^2]^{1/2}},
\end{equation}
for a closed universe, $\Omega_k < 0$, and:
\begin{equation}
\theta_s^{-1} \approx {I \over 2c_s~a_{dec}^{1/2}}
\left(\Omega_m\right)^{1/2};\;\; 
I=\int_{a_{dec}}^1 {c~ da\over 
 [\Omega_m a+\Omega_\Lambda a^4]^{1/2}}.
\end{equation}
for a flat universe, $\Omega_k = 0$. \footnote{These approximate expressions 
were derived neglecting the time variation of the sound velocity and 
the contribution of relativistic species. 
Both effects, which may be substantial for models with low matter content, 
were taken into account when calculating the curves shown in Fig. 2.}

This very simplified discussion gives an intuitive idea of how the main
CMB observables, namely the position and height of peaks in the $C_\ell$'s,
are affected by cosmological parameters in inflationary adiabatic models. 
For a given primordial power
spectrum, the anisotropy pattern is defined by the baryon content
and dark matter content, $\Omega_b h^2$ and $\Omega_{cdm}h^2$, 
which basically affect the amplitude of fluctuations
at different physical scales and fix the relative heights of the peaks. 
This structure is mapped into different angular
scales depending on a combination of $\Omega_m$, $\Omega_\Lambda$ and 
$\Omega_k$, which affect the position of the peaks. These effects
are illustrated in Figure 1. 
This also shows that the parameters actually enter in 
the power spectrum in a combined way, leading in some cases to almost
exact degeneracies: a dramatic example of this behavior is shown in Figure 2.
The problem of parameter degeneracies has been explored in great detail in
\cite{balbi:eb99}.

\begin{figure}[t!] 
\centerline{\epsfig{file=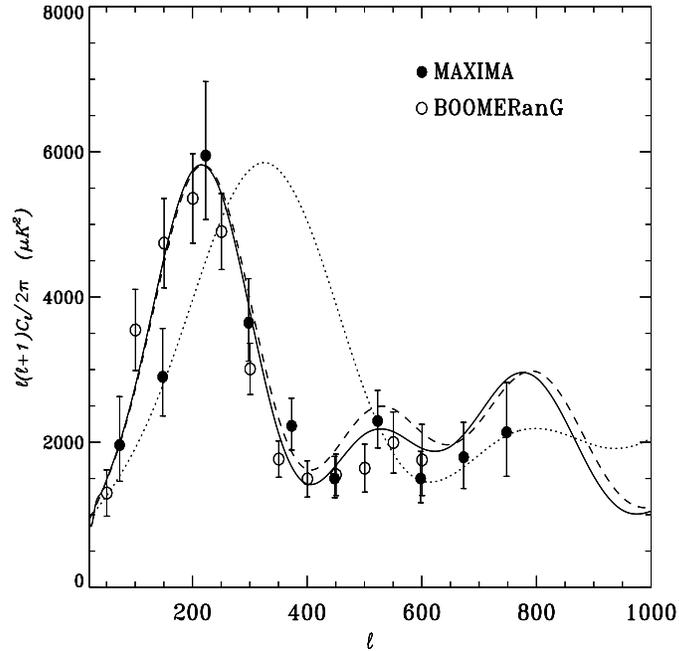,height=3.5in,width=3.5in}}
\vspace{10pt}
\caption{The angular power spectrum of the CMB for inflationary adiabatic
models with scale invariant primordial fluctuations (n=1). The three curves
show the effect of varying the value of the parameters {\bf p}=
($\Omega=\Omega_m+\Omega_\Lambda$, $\Omega_b h^2$, $\Omega_{cdm}h^2$, $\Omega_\Lambda$) 
is shown in the three curves. The
solid curve has {\bf p}=(1, 0.03, 0.2 0.54). 
If $\Omega_b h^2$ and $\Omega_{cdm}h^2$ 
are kept fixed while varying $\Omega_m$ and $\Omega_\Lambda$, the peaks
are simply shifted (dotted curve, with {\bf p}=(0.46, 0.03, 0.2, 0)). 
When we fix $\Omega_m$ and $\Omega_\Lambda$ and 
we vary the baryon and cold dark matter
content, the positions are basically unchanged, and the ratio of peaks
height changes (dashed curve, with {\bf p}=(1, 0.025, 0.17, 0.54)). 
The points are the bandpower measurements from BOOMERanG and MAXIMA, 
re-scaled within their 1$\sigma$ calibration uncertainty.}
\end{figure}

\begin{figure}[t!] 
\centerline{\epsfig{file=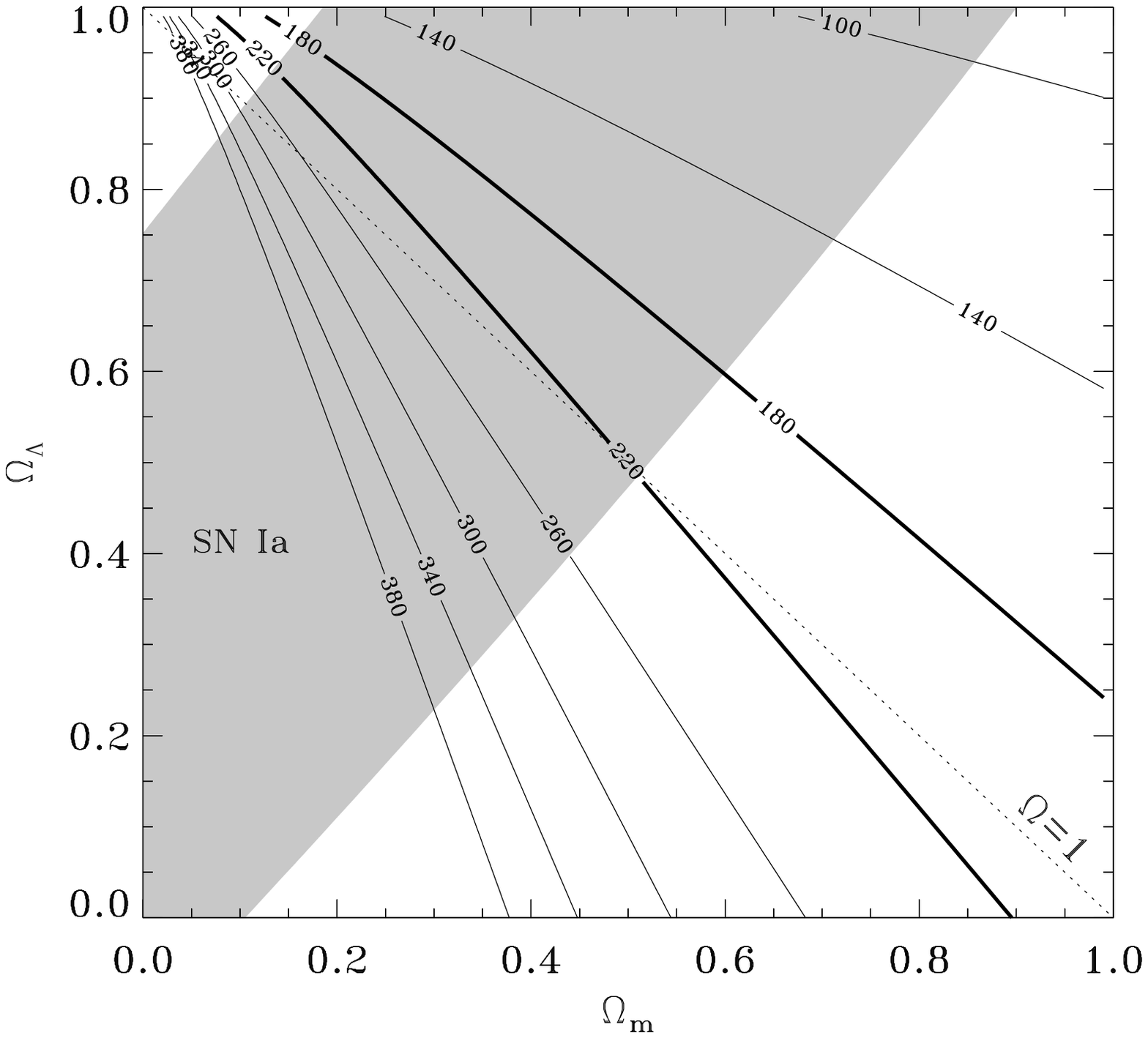,height=4.5in,width=4.5in}}
\vspace{10pt}
\caption{Degeneracy lines in the $\Omega_m$--$\Omega_\Lambda$ plane, assuming
$h=0.7$ and $\Omega_b h^2=0.03$. The lines correspond to models 
having approximately the same angular scale of sound horizon at decoupling, 
and are labeled by
the approximate position of the first peak in the CMB angular power spectrum. 
The ticker lines give a reference interval corresponding to $180\leq 
\ell_{peak} \leq 220$. Flat models lie on the diagonal dotted line
($\Omega=1$). The shaded curve represents the 95 \% confidence levels from 
high redshift type Ia supernovae.}
\label{balbi:fig2}
\end{figure}

\section*{Cosmological Parameter Estimation from CMB Measurements}

As we saw in the previous Section, a given cosmological model is 
specified by a set of cosmological parameters: 
\begin{equation}
{\bf p}=\{p_1,p_2,\cdots,p_n\}$$ 
\end{equation}
entering in the calculation of the theoretical power spectrum, 
$\bar{C_\ell}$.
A CMB experiment measures the temperature fluctuation 
of the CMB in different 
directions on the sky. These data are used to build a minimum variance
map of the CMB. From the map, maximum likelihood estimates of the 
power spectrum are extracted
as a set of average bandpower measurements over an interval $b$ in $\ell$: 
\begin{equation}
{\cal C}_b\equiv\sum_{\ell\in b}S_\ell C_\ell.
\end{equation}
where the shape function $S_\ell$ is usually assumed to be 
$S_\ell\propto \ell(\ell+1)$.  

The best estimate of the cosmological parameters from a set of measured
bandpowers
can be derived in a Bayesian sense by maximizing the likelihood function:
\begin{equation}
{\cal L}({\bf p})\propto {\cal P}({\cal C}_b\vert {\bf p})
{\cal P}({\bf p}\vert {\rm prior})
\end{equation}
We should note that the likelihood is not a Gaussian function of 
the ${\cal C}_b$'s. One way to estimate how the likelihood depends
on the ${\cal C}_b$'s is to use the ansatz described
in \cite{balbi:bjk00} where an 
additional quantity $x_b$ related to the noise of the experiment is 
needed in order to fully characterize the likelihood. 
This has been shown to work well 
when compared to a brute force exact calculation of
the likelihood. 

In a Bayesian framework, getting the {\em a posteriori}
likelihood of one (or more) parameters
out of the full likelihood, involves an integration ({\em marginalization}) of
the likelihood over the unwanted ({\em nuisance}) parameters. 
The full likelihood has to be weighted by the {\em a priori} 
likelihoods of the nuisance parameters:
\begin{equation}
{\cal L}(p_i,\cdots,p_j)=\int dp_l\cdots dp_m 
{\cal L}(p_i,\cdots,p_j\vert p_l,\cdots,p_m){\cal L}(p_l)\cdots{\cal L}(p_m)
\end{equation}
This complicates the analysis considerably, making the problem strongly
non-local. The overall shape of the likelihood influences the result
on subsets of parameters, and we then need to characterize the 
likelihood over the entire parameter space even if we are just interested 
in a subset of parameters. The problem is exacerbated by the presence
of correlations in the parameter space. In addition, we need a knowledge 
(or a reasonable guess) of the {\em a priori} likelihood of the 
nuisance parameters. This can be chosen to be uniform 
(in the absence of previous knowledge) or can be derived from
other observations.
We also stress the fact that the extent of the parameter space 
effectively acts as a top-hat
prior, cutting out {\em a priori} some values of the parameters.
In general
one should carefully explore how changing the
priors on some parameters affects the results for the others.
This is particularly important when the results for some parameters conflict 
with those coming from other observations.
The issues related to the  use of different
priors have been thoroughly explored
in the analysis of the latest CMB datasets (see, e.g. \cite{balbi:balbi00}
\cite{balbi:lange00} and \cite{balbi:jaffe00}).

\section*{Current Constraints}

Over the past decade, many experiments have gathered data on CMB 
anisotropy, which, taken collectively, gave an indication of the shape
of the angular power spectrum. Such data were used by a number of authors
to set constraints on the parameters of the inflationary adiabatic model 
(see, e.g. \cite{balbi:tegmark}, and references therein). 
The quality of CMB data has considerably improved after the recent
BOOMERanG\footnote{http://oberon.roma1.infn.it/boomerang}
\cite{balbi:debe00} and 
MAXIMA\footnote{http://cfpa.berkeley.edu/maxima} \cite{balbi:hanany00} 
balloon-borne missions. BOOMERanG estimated the 
power spectrum in 12 bins, over the range $26\leq\ell\leq 625$,
from a map of a 1800 square degrees patch of the southern sky; 
MAXIMA estimated the spectrum in 10 bins, 
over the range $36\leq\ell\leq 785$, from a map of a 124 square degrees
patch of the northern sky. 
The two measurements are in remarkable agreement and show unambiguously 
the presence of a sharp peak in the region $180 \leq \ell \leq 220$ 
(see Figure 1). This is, in itself, a strong evidence in favor
of inflationary adiabatic models; 
alternative theories either predict a
broader peak at higher $\ell$ or a broad shelf at $\ell<200$ (see
e.g. \cite{balbi:knox}).

\begin{table}
\caption{Current constraints (68\% c.l.) from CMB data only.}
\begin{tabular}{lcccc}\medskip
Dataset & $\Omega$ & $\Omega_b h^2$ & $\Omega_{cdm}h^2$ & $n$ \\
\tableline
BOOMERanG + COBE & $1.15^{0.10}_{0.09}$ & $0.036^{0.006}_{0.005}$ & 
$0.24^{0.08}_{0.09}$ & $1.04^{0.10}_{0.09}$ \medskip  \\ 
MAXIMA + COBE & $1.00^{0.07}_{0.15}$ & $0.030^{0.005}_{0.005}$ & 
$0.20^{0.10}_{0.05}$ & $1.08^{0.05}_{0.05}$ \medskip \\
MAXIMA + BOOMERanG + COBE & $1.11^{0.07}_{0.07}$ & $0.032^{0.005}_{0.005}$ & 
$0.14^{0.06}_{0.05}$ & $1.01^{0.09}_{0.08}$ \medskip \\
\end{tabular}
\end{table}

\begin{table}
\caption{Current constraints (68\% c.l.)
from CMB data plus prior information from high redshift type Ia supernovae}
\begin{tabular}{lcc}\medskip
Dataset & $\Omega_\Lambda$ & $\Omega_m$ \\
\tableline
BOOMERanG + COBE + SNe Ia & $0.72^{0.05}_{0.04}$ & 
$0.37^{0.07}_{0.07}$ \medskip \\
MAXIMA + COBE + SNe Ia & $0.60^{0.07}_{0.07}$ & $0.37^{0.07}_{0.07}$ \medskip \\
MAXIMA + BOOMERanG + COBE + SNe Ia & $0.75^{0.06}_{0.07}$ & $0.35^{0.07}_{0.07}$ \medskip \\
\end{tabular}
\end{table}

The BOOMERanG and MAXIMA datasets (in combination with COBE) 
were used independently to set constraints on 
a seven-dimensional space of cosmological parameters \cite{balbi:balbi00} 
\cite{balbi:lange00};
a joint analysis was also carried on \cite{balbi:jaffe00}.
Using no prior information (except the constraint that the universe is older
than 10 Gyr and that the Hubble constant is $0.45\leq h \leq0.90$) the
analyses performed in \cite{balbi:balbi00} \cite{balbi:lange00} and
\cite{balbi:jaffe00} found the results reported in Table 1.
These CMB data alone are already giving a consistent picture, which is in 
agreement with the basic predictions of inflation: the universe is nearly
flat, and the primordial fluctuations have a nearly scale-invariant 
power spectrum. Moreover, there is indication of a substantial contribution 
from non-baryonic matter. It may also be observed that the best estimate of
the baryon density from CMB shows some tension with the value from big bang
nucleosynthesis \cite{balbi:bbn} $\Omega_b h^2=0.019\pm 0.002$. Some
implications of this discrepancy have been explored in \cite{balbi:esposito}. 

As we saw in the first Section, models with different values of 
$\Omega_m$ and $\Omega_\Lambda$ may result in the same angular 
scale of sound horizon at decoupling. As a consequence, it is hard for
CMB data alone to separately determine $\Omega_m$ and $\Omega_\Lambda$.
This is clear from Figure 2 where we plot lines 
in the $\Omega_m$---$\Omega_\Lambda$ plane corresponding to the
same angular scale of sound horizon at decoupling. To break this degeneracy,
prior information from different datasets can be used. The results obtained
by \cite{balbi:balbi00} \cite{balbi:lange00} and
\cite{balbi:jaffe00} when combining 
the CMB data with constraints in the $\Omega_m$---$\Omega_\Lambda$ plane
coming from observations of high-redshift type Ia supernovae 
\cite{balbi:perlm} \cite{balbi:riess} are shown in Table 2.
These results make a very strong case for the existence of a substantial
contribution from some form of unknown negative-pressure
component (named ``dark energy'' in the recent literature).
Attempts to investigate the nature of this component using the CMB, while
assuming strong priors on the other parameters, have been made in
\cite{balbi:quint}.

Finally, it is remarkable that, as shown in \cite{balbi:jaffe00},
when completely independent priors from large
scale structure observations are used instead of those from supernovae, 
totally consistent results are obtained.

The CMB is already proving very powerful in improving our knowledge of 
cosmological parameters. Increasingly accurate measurements of the power
spectrum will come in the next few years by satellite missions such as
MAP\footnote{http://map.gsfc.nasa.gov} and 
Planck\footnote{http://astro.estec.esa.nl/Planck/}, 
which will further strengthen our understanding of the nature of the universe. 

\section*{Acknowledgments}
I wish to thank the MAXIMA and BOOMERanG collaborations.
I am also grateful to
Saurabh Jha and the High-Z Supernova Search Team for 
providing the likelihood function from supenovae measurements, 
and to Giancarlo de Gasperis, Pedro Ferreira, Paolo Natoli and Nicola Vittorio 
for useful discussions.


\begin{references}
\bibitem{balbi:smoot92} Smoot, G.F., et al., {\it ApJ}, {\bf 396}, L1 (1992).
\bibitem{balbi:debe00} de Bernardis, P., et al., {\it Nature}, {\bf 404}, 
955 (2000).
\bibitem{balbi:hanany00} Hanany, S., et al., {\it ApJL}, accepted, 
astro-ph/0005123 (2000).
\bibitem{balbi:balbi00} Balbi, A., et al., {\it ApJL}, accepted,
astro-ph/0005124 (2000).
\bibitem{balbi:lange00} Lange, A.E., et al., {\it Phys. Rev. D}, submitted, 
astro-ph/0005004 (2000).
\bibitem{balbi:tegmark00} Tegmark, M., \& Zaldarriaga, M., 
{\it Phys. Rev. Lett.}, {\bf 85}, 2240 (2000).
\bibitem{balbi:jaffe00} Jaffe, A., et al., {\it Phys. Rev. Lett.}, submitted, 
astro-ph/0007333 (2000).
\bibitem{balbi:schwarz00} Schwarz, D.J., Martin, J., Riazuelo, A., this volume,
astro-ph/0010453 (2000).
\bibitem{balbi:cmbfast} Seljak, U.  \& Zaldarriaga, M., 
{\it ApJ}, {\bf 469}, 437 (1996).
\bibitem{balbi:hu} Hu, W., Sugiyama, N. \& Silk, J., {\it Nature}, 
{\bf 386}, 37-43 (1997).
\bibitem{balbi:eb99} Efstathiou G. \& Bond, J.R., {\it MNRAS}, {\bf 304}, 
75 (1999)
\bibitem{balbi:bjk00} Bond, J.R., Jaffe, A.H., \& Knox, L., {\it ApJ}, 
{\bf 533}, 19 (2000).
\bibitem{balbi:tegmark} Tegmark, M., \& Zaldarriaga, M., {\it ApJ} in press,
astro-ph/0002091 (2000).
\bibitem{balbi:knox} Knox, L., \& Page, L. {\it Phys. Rev. Lett.}, submitted,
astro-ph/0002162 (1999).
\bibitem{balbi:bbn} Burles, S., Nollett, K.M., Truran, J. N., \& Turner, M. S.
{\it Phys. Rev. Lett.}, {\bf 82}, 4176 (1999).
\bibitem{balbi:esposito} Esposito, S., et al., astro-ph/0007419 (2000).
\bibitem{balbi:perlm} Perlmutter, S., et al., {\it ApJ}, {\bf 517}, 
565 (1999).
\bibitem{balbi:riess} Riess, A.G., et al., {\it AJ}, {\bf 116}, 1009 (1998).
\bibitem{balbi:quint} Balbi, A., Baccigalupi, C., Matarrese, S., Perrotta, F.
\& Vittorio, N., submitted to {\it ApJL}, astro-ph/0009432 (2000).
\end{references}
\end{document}